\documentclass[preprint,12pt]{elsarticle}

\usepackage{epsfig}
\usepackage{lineno}
\usepackage{amssymb}
\usepackage[figuresright]{rotating}
\usepackage{rotating}
\usepackage{subfig}
\usepackage{graphicx}
\usepackage{color}
\usepackage{pdflscape}

\journal{Nucl. Instrum. Meth. Phys. Res. A}

\begin{document}

\begin{frontmatter}

\title{Charge-to-heat transducers exploiting the Neganov-Trofimov-Luke effect for light detection in rare-event searches}

\author[CSNSM]{V.~Novati}
\author[CSNSM]{L.~Berg\'e}
\author[CSNSM]{L.~Dumoulin}
\author[CSNSM,DISAT]{A.~Giuliani}
\author[CSNSM]{M.~Mancuso}
\author[CSNSM]{P.~de~Marcillac}
\author[CSNSM]{S.~Marnieros}
\author[CSNSM]{E.~Olivieri\corref{cor3}}
\cortext[cor3]{corresponding author}
\ead{emiliano.olivieri@csnsm.in2p3.fr}
\author[CSNSM,KINR]{D.V.~Poda}
\author[CSNSM]{M.~Tenconi}
\author[CEA-IRFU]{A.S.~Zolotarova}

\address[CSNSM]{CSNSM, Univ. Paris-Sud, CNRS/IN2P3, Universit\'e Paris-Saclay, 91405 Orsay, France}
\address[DISAT]{DISAT, Universit\`a dell'Insubria, 22100 Como, Italy}
\address[CEA-IRFU]{IRFU, CEA, Universit\'{e} Paris-Saclay, F-91191 Gif-sur-Yvette, France}
\address[KINR]{Institute for Nuclear Research, 03028 Kyiv, Ukraine}

\begin{abstract}

In this work we present how to fabricate large-area (15~cm$^2$), ultra-low threshold germanium bolometric photo-detectors and how to operate them to detect few (optical) photons. These detectors work at temperatures as low as few tens of mK and exploit the Neganov-Trofimov-Luke (NTL) effect. They are operated as charge-to-heat transducers: the heat  signal is linearly increased by simply changing a voltage bias applied to special metal electrodes, fabricated onto the germanium absorber, and read by a (NTD-Ge) thermal sensor. We fabricated a batch of five prototypes and ran them in different facilities with dilution refrigerators. We carefully studied how  impinging spurious infrared radiation impacts the detector performances,  by shining infrared photons via optical-fiber-guided LED signals, in a controlled manner, into the bolometers. We hence demonstrated how the radiation-tightness of the test environment tremendously enhances the detector performances, allowing to set  electrode voltage bias up to 90 volts without any leakage current and signal-to-noise gain as large as a factor 12 (for visible photons). As consequence, for the first time we could operate large-area NTD-Ge-sensor-equipped NTL bolometric photo-detectors capable to reach sub 10-eV baseline noise (RMS). Such detectors open new frontiers for rare-event search experiments based on low light yield Ge-NTD equipped scintillating bolometers, such the CUPID neutrinoless double-beta decay experiment.
\end{abstract}


\end{frontmatter}


\section{Introduction}

Bolometric light detectors are nowadays employed in several cryogenic experiments searching for rare events, as direct detection of dark matter (CRESST~\cite{Angloher:2016a}, COSINUS~\cite{Angloher:2016}) and searches for neutrinoless double-beta decay (AMoRE~\cite{Alenkov:2018}, LUCIFER/CUPID-0~\cite{Azzolini:2018}, LUMINEU~\cite{Armengaud:2017} and its follow-up CUPID-Mo~\cite{Poda:2017}). 
They are coupled to the main scintillating crystals which contain the nuclear targets for the dark-matter particles or the nuclei that can undergo neutrinoless double-beta decay. The main crystal is operated as a bolometer and the simultaneous detection of heat and light signals associated to the same event can provide particle identification and consequently an active background rejection, as proposed in~\cite{Bobin:1997,Alessandrello:1998,Meunier:1999,Pirro:2006t} and successfully performed (see e.g. in~\cite{Pirro:2017,Poda:2017a,Bellini:2018}). In particular, composite heat-and-light detectors allow to control dominant background events such as nuclear recoils in dark-matter searches and alpha particles in double-beta-decay experiments.

In dark-matter searches, a high-performance light detector is required to lower the energy threshold~\cite{Stark:2005} and to identify recoils of different-mass nuclei~\cite{Angloher:2012}. In double-beta decay searches, high-sensitivity light detectors are needed either to detect the feeble Cherenkov light emitted by poorly-scintillating crystals~\cite{Fatis:2010} (this is the case of the promising compound TeO$_2$~\cite{Brofferio:2018,Tretyak:2002,Alduino:2017,Wang:2015a,Wang:2015b}), or to help in pile-up rejection (as in $^{100}$Mo-enriched bolometers, ~\cite{Chernyak:2012,Chernyak:2014,Chernyak:2016}). The pile-up rejection capability is useful also to perform precision calorimetric measurements of rare-$\beta$-decay spectral shapes (as those of $^{113}$Cd and $^{115}$In) which can be used to scrutinize the value of the axial-vector coupling constant~\cite{Tretyak:2017,Suhonen:2017,Leder:2018}.
\newline
\\
Neganov-Trofimov-Luke (NTL) effect~\cite{Neganov:1985,Luke:1988} can be exploited in high purity semiconductor-based bolometers for lowering the detection threshold and enhancing the signal-to-noise-ratio.

In case of an ionizing particle of primary energy $E_0$ interacting in a semiconductor absorber, an extra heat energy is produce if charge carriers, created by the particle interaction, are drifted by an electric field.
The total heat sensed by the bolometer is:
\begin{equation}
E_{tot} = E_0 \left( 1 + \frac{q \cdot V_{el} \cdot \eta }{\epsilon }  \right) = E_0 \cdot G_{NTL},
\end{equation}
where $\epsilon$ is the average energy required to generate an electron-hole pair, $q$ is the elementary charge and $V_{el}$ is the charge collecting potential between electrodes (Fig. \ref{fig:NTLLD}, bottom panel), deposited on the germanium absorber to set a drifting electric field across this latter; $\eta$ is an amplification efficiency which accounts for an incomplete gain due to charge trapping or other losses (in an ideal case $\eta$ = 1).

For $V_{el}>>\epsilon/q$ (NTL regime), the heat energy $E_{tot}$ is mainly due to the NTL effect; therefore, a bolometer operated under this condition behaves as a voltage-controlled charge-to-heat amplifier. 
The NTL signal amplification is observed in EDELWEISS~\cite{Hehn:2016} and CDMS~\cite{Agnese:2016} dark-matter search experiments, which employ high purity germanium and silicon ionization-and-heat composite bolometers.

To-date, several technologies of NTL bolometers for the detection of photons have been developed and used for the aforementioned applications. They can be grouped according to the absorber material and the temperature sensor as follows: silicon absorbers equipped with TES (Transition-Edge Sensor) thermometers~\cite{Stark:2005,Isaila:2006,Isaila:2012,Willers:2015,Defay:2016}; silicon absorber with NTD (Neutron-Transmutation-Doped) germanium thermistors~\cite{Biassoni:2015,Gironi:2016}; and germanium absorbers read-out by NTD-Ge thermistors~\cite{Pattavina:2016,Artusa:2017,Berge:2017}.\\

In this work we report on the fabrication method of NTL, NTD-Ge equipped germanium bolometers and present the development, characterization and performance of five of them. Additional information and measurements done with early NTL bolometric light detector prototypes can be found in~\cite{Tenconi:2015,Mancuso:2016,Novati:2018}. 
\section{Detector fabrication}

The development of NTL-effect-assisted cryogenic light detectors has been carried out at CSNSM laboratory (Orsay, France).
The detector absorbers are done by electronic-grade germanium wafers (impurity level of the order of 10$^{11}$/cm$^3$) of 44~mm diameter and 0.175~mm thickness, supplied by UMICORE. 
Wafers are bombarded with argon ions to remove the germanium oxide at the surfaces and improve the adherence of any subsequent structure deposition onto the surfaces. 
A 50-nm-thick hydrogenated-amorphous germanium layer is evaporated, as discussed in~\cite{Shutt:2000}. 
Five 100-nm-thick, 3.8-mm-pitch annular concentric aluminium electrodes are then deposited. Finally the wafers are coated by a 70-nm-thick SiO layer, to enhance the light absorbtion\footnote{The enhancement of visible-wavelength photons absorption was initially found to be 35\%~\cite{Mancuso:2014}.}.
\begin{figure}[htb]
  \begin{center}
  \begin{minipage}[t]{0.75\textwidth}
	\includegraphics[width=\textwidth]{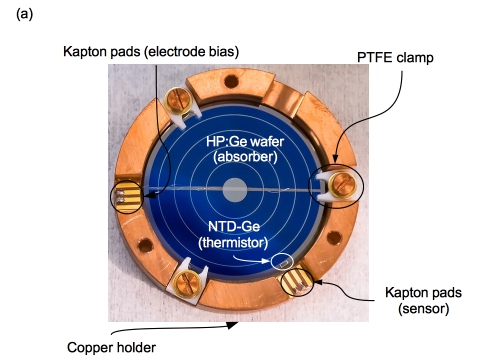}
  \end{minipage}
  \vspace{2mm}\\
  \begin{minipage}[t]{0.75\textwidth}
	\includegraphics[width=\textwidth]{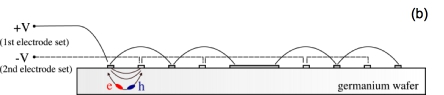}
  \end{minipage}
 \caption{(a) A picture of a NTL-assisted light detector (NTLLD1). A thin region (grey strip crossing the annular electrodes) without any SiO coating is visible; it is used to connect via ultrasonic bonding the different annular electrodes (light circles) and create two sets of bias electrodes.
(b) Sketch of the electrical connection between annular electrodes. An electrical potential $\Delta V= V_+ - V_-$ can be applied to the two sets of bias electrodes, to drift electrons (e) and holes (h) as depicted in the scheme, along the electric field lines (black, solid).}
\label{fig:NTLLD}
  \end{center}
\end{figure}
The germanium absorbers are mounted in copper holders and kept by three PTFE clamps; absorbers are instrumented with a 5~mg NTD-Ge thermistors. The electrical contacts between the detectors and the cryostat cabling is ensured by Kapton-insulated Au-coated copper pads, glued on the copper casing. The NTD-Ge thermistors are electrically connected to the aforementioned pads through 25~$\mu$m diameter, $\sim$10~mm long gold bonding wires. These latter act also as thermal link between the absorbers and the thermal bath. The aluminum electrodes are connected by 25~$\mu$m diameter aluminum bonding wires to form two separated sets of bias electrodes; a voltage can be applied (Fig. \ref{fig:NTLLD}(b)) to set a charge carrier drifting field within the semiconductor absorber. Five NTL light detectors (from NTLLD0 to NTLLD4) have been fabricated according to the above-presented scheme. For two of them, the process was slightly changed: NTLLD0 did not have any SiO coating and NTLLD4 was equipped with an NTD-Ge of smaller mass ($\sim$2~mg) in order to enhance the sensitivity. Fig.~\ref{fig:NTLLD}(a) shows a picture of NTLLD1 detector.
%
%
\section{Detector operation}
\subsection{Equipment and conditions of low-temperature tests}
The NTL light detectors have been first operated aboveground, in different environments, i.e. two dry and one wet dilution refrigerators, at the CSNSM laboratory \cite{Mancuso:2014a,Mancuso:2016}. Some of them have been afterwards operated underground at LNGS (Laboratory Nazionali del Grans Sasso, Italy) and at LSM (Laboratoire Souterrain de Modane, France) where the  CUPID~R\&D and EDELWEISS-III experiments are located, respectively (a description of both cryogenic facilities can be found e.g. in~\cite{Mancuso:2016,Armengaud:2017}). In order to reduce the noise due to vibrations generated by the pulse-tube of the dry refrigerators~\cite{Olivieri:2017}, spring-loaded mechanical decoupling systems (see e.g., in~\cite{Pirro:2006,Lee:2017,Armengaud:2017,Maisonobe:2018}) have been used.  

Only in a few measurements the detectors under study were investigated with visible light photons emitted by scintillating crystals; most of the tests have been carried out with near-infrared photons, provided by an infrared LED setup located at room temperature and guiding the photons via 0.2-mm diameter, plastic optical fibers down to the light detectors. 0.85~$\mu$m wavelength (Honeywell HFE4050) and 0.95~$\mu$m wavelength (Osram LD271) photon packages (bursts) were delivered, capable to provide total energies ranging from a few eV up to few MeV in a single burst, by simply changing the driving parameters (electrical pulse width and amplitude) of the LEDs.
The LEDs were also used to charge-reset the germanium absorbers~\cite{Olivieri:2009}.

Most of the measurements have been performed at temperatures as low as 15~mK. A room-temperature, DC electronics with a 675~Hz lowpass cut-off frequency ~\cite{Alessandrello:2000} has been used to shape the thermistor signals, which have been sampled at 5$\div$10~kHz to correctly reconstruct the signal shape.
\subsection{Optimal working point settings}
\begin{figure}[ht] 
\centering
    \includegraphics[width=0.75\textwidth]{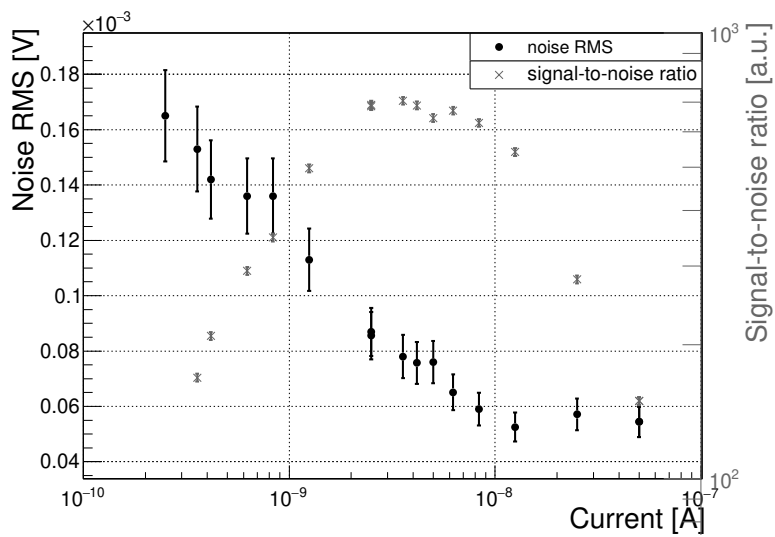}	
  \caption{A signal-to-noise ratio (crosses) of the NTLLD2 detector as a function of the bias. The RMS noise (points) is also shown. For this detector and set-up, the maximal SNR is observed for a bias of 3~nA.}
	\label{fig:WorkingPoint}
\end{figure}

The NTD-Ge thermistor bias for the tested detectors has been chosen by stabilizing the temperature of the mixing chamber and searching for the maximal signal-to-noise ratio (SNR). To this end, we delivered constant LED pulses (signal) and recorded the detector RMS baseline noise, for different NTD-Ge sensor biases (Fig. \ref{fig:WorkingPoint}).
\subsection{Calibration}
\begin{figure}
\subfloat[\label{Calibration:a}][Energy spectrum of an X-ray $^{55}$Fe source, irradiating the NTLLD1 detector.]
{\includegraphics[width=.45\linewidth]{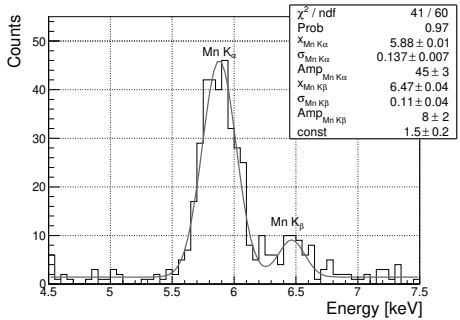}}\hfill
\subfloat[\label{Calibration:b}][Muon signal distribution recorded with NTLLD4. The maximum of the distribution corresponds to about 100~keV (see text).]
{\includegraphics[width=.45\linewidth]{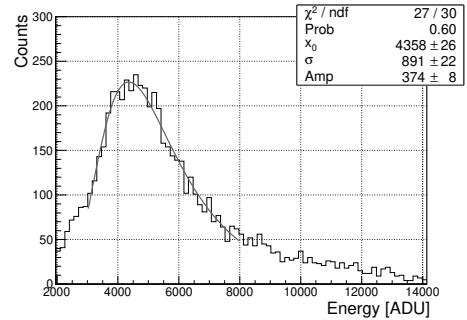}}\par
\subfloat[\label{Calibration:c}][LED pulses (bursts) of different intensities delivered to the detector (NTLLD2), which records signal of x$_0$ amplitude (uncalibrated). Each peak has (x$_0$)$_i$ mean and $\sigma^2_i$ width.]
{\includegraphics[width=.45\linewidth]{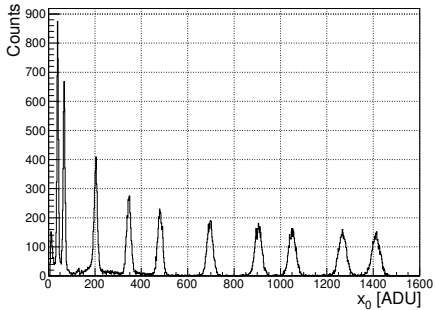}}\hfill
\subfloat[\label{Calibration:d}][Squared peak width $\sigma^2_i$ as a function of (x$_0$)$_i$. The number of photons impinging the detector is given by $a\cdot$x$_0$. The detector response (calibration) is derived by knowing the average wavelength of the LED photons.]
{\includegraphics[width=.45\linewidth]{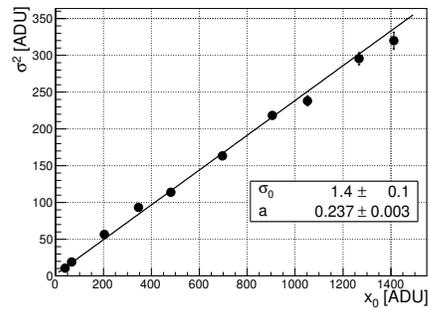}}
 \caption{Illustration of different detector calibration methods.}
 \label{fig:Calibration}
 \end{figure}
Several approaches have been used to calibrate the NTL bolometers, such as: X-rays (e.g. a $^{55}$Fe source, X-ray fluorescence induced by a high-intensity $\gamma$ source), scintillation, LED photon bursts and photon statistics, environmental muons.

A $^{55}$Fe source was often used to irradiate the detector absorber. A typical spectrum obtained is shown in Fig.~\ref{fig:Calibration}(a). 

As an alternative method, an external high-activity gamma source can be used to induce the X-ray fluorescence of the materials directly surrounding the detector~\cite{Berge:2017}. One can also exploit the cosmic rays in an aboveground facility, by recording the bolometer signal distribution of muons crossing the germanium wafer. 
The muon energy loss probability is well-described by the Landau distribution~\cite{Landau:1944} and is illustrated in Fig.~\ref{fig:Calibration}(b). 
The most probable muon-induced energy release has been evaluated by Geant4-based Monte Carlo simulation\footnote{The simulation has been performed in a very simplified approach, assuming a cosmic muon angular distribution proportional to $\cos^2(\Theta_Z)$, where ${\Theta}_Z$ is the zenith angle. The generated particles are $\mu^+$'s with 3~GeV energy.} of the order of 100~keV (for a 0.175~mm thick germanium absorber).

The detector calibration can also be performed by using LED photon bursts and photon statistics, as successfully demonstrated in ~\cite{Isaila:2012}, and reported in Fig.~\ref{fig:Calibration}(c),(d). 

In NTL regime, we could not use the $^{55}$Fe X-ray lines (5.9 and 6.5 keV) to calibrate our detector energy response, since those lines were broad and washed-out.
This behaviour is manly due to: (1) e-h charge recombination in the primary plasma, created when the photon interacts within the germanium absorber. X-rays release all their energy in a well-defined point of the absorber, creating a dense plasma of electron-hole pairs. The electric field set in the germanium absorber via the bias electrodes separate and drift only the external charges (plasma erosion) whereas the internal e-h pairs eventually recombine. This leads to an incomplete charge collection and hence a broadening of the detector signal; (2) e-h trapping due to defects, impurities and surface effects. Again, the e-h charges created by an ionising particle are trapped (while drifted toward the bias/collecting electrodes) and the detector experiences an incomplete charge collection.

For peak provided via LED photon bursts (where photons interact in the germanium simultaneously but over a large area) and for muon interactions this broadening is not observed. This is probably due to the fact that the e-h charge density is low and the charges are separated and drifted before recombining.
Therefore only these two latter techniques could be used for the calibration of the energy detector response in the NTL regime.

When a light detector is coupled to a source of scintillation and/or Cherenkov radiation, the registered light (initially calibrated e.g. by X-rays at 0~V electrode bias) can also be exploited ~\cite{Artusa:2017}. 
It is worth noting that the calibration of a detector operated in the NTL mode strongly depends on the source used, since the quantum efficiency $\epsilon$ depends on the particle and wavelength (we will see later in the text). 
In our application, the main purpose of NTL light bolometers is the detection of visible wavelength photons. Therefore the scintillation light is the most pertinent source of calibration.
\section{Detector performance}
\subsection{Photo-current noise}
\label{sec:Photonoise}
As observed in the very early investigations of NTL detectors~\cite{Stark:2005,Isaila:2006,Isaila:2012}, the performances (\emph{i.e.} baseline noise and resolution) of these devices are strongly degraded if they receive a spurious photon flux (for example, coming from high temperature black-bodies). In particular, when a voltage bias is set on the bias electrodes, a power proportional to the photon flux times the voltage is dissipated; consequently the detector warms up and settles to a different working point. Before experiencing this heating, the detector shows a baseline excess noise.
\begin{figure}[h!]
\subfloat[\label{Photocurrent:a}][Gain and NTD-Ge resistance, as a function of the electrode voltage bias and different LED photon flux intensity.]
{\includegraphics[width=.45\linewidth]{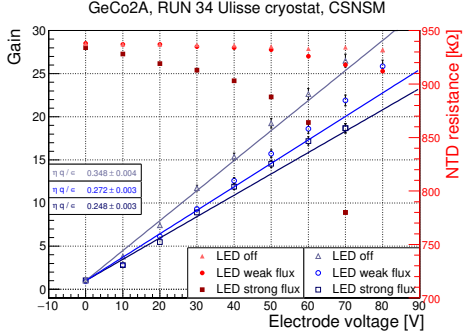}}\hfill
\subfloat[\label{Photocurrent:b}][Signal-to-noise ratio as a function of the electrode voltage bias, for different photon fluxes.]
{\includegraphics[width=.45\linewidth]{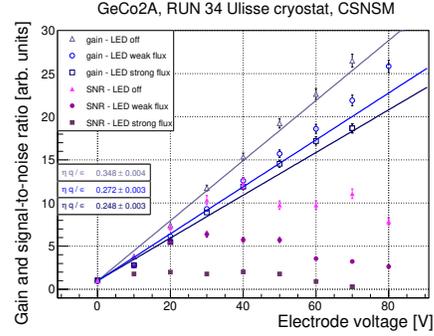}}\par
\subfloat[\label{Photocurrent:c}][Noise spectra of NTLLD4 detector signal acquired for different electrode voltage bias, without photon flux.]
{\includegraphics[width=.45\linewidth]{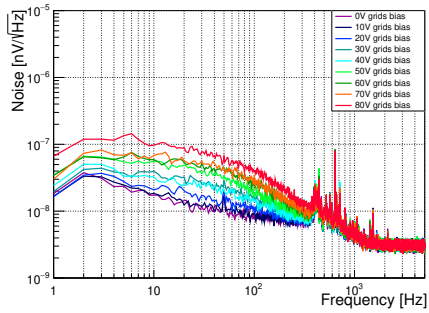}}\hfill
\subfloat[\label{Photocurrent:d}][Noise spectra of NTLLD4 detector signal, acquired for different electrode voltage bias and for different photon flux generated via the LED.]
{\includegraphics[width=.45\linewidth]{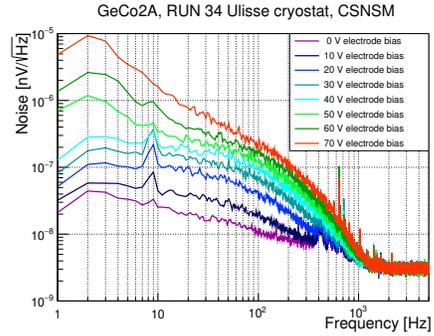}}
 \caption{(\textit{Color online}) NTL detector behaviour studies performed to inspect the impact of a spurious photon flux impinging the germanium (semiconductor) absorber.}
 \label{fig:Photocurrent}
 \end{figure}

To understand the aforementioned behaviour, we shined the detector with a constant photon flux in a controlled manner, by using an infrared LED. We monitored the NTL gain, the signal-to-noise ratio and the NTD-Ge resistance as a function of the electrodes bias, for different LED photon flux intensities. Fig.~\ref{fig:Photocurrent} gathers the results obtained with the NTLLD4 detector. The NTL gain still increases with respect to the electrode bias but the gain factor (slope) decreases, mainly due to a sensitivity degradation which is caused by the warming up of the working point (Fig.~\ref{fig:Photocurrent}(a)).
Moreover, the photo-current injection drastically affects the signal-to-noise ratio; at 40~V a reduction of this latter as high as 50\% (80\%) for the weak (strong) LED-driven photon flux is observed (Fig.~\ref{fig:Photocurrent} (c) and Fig.~\ref{fig:Photocurrent}(d)).

Therefore, the performance of NTL bolometers can be drastically improved by carefully shielding against spurious radiation, making the detector photon-tight with respect to environmental photons.
\subsection{Performance in NTL regime}
\begin{figure}[h!]
\subfloat[\label{Gain_SNR:a}][NTLLD0 gain and signal-to-noise ratio.]
{\includegraphics[width=.45\linewidth]{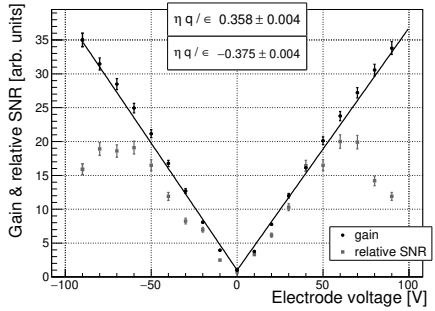}}\hfill
\subfloat[\label{Gain_SNR:b}][NTLLD1 gain and signal-to-noise ratio.]
{\includegraphics[width=.45\linewidth]{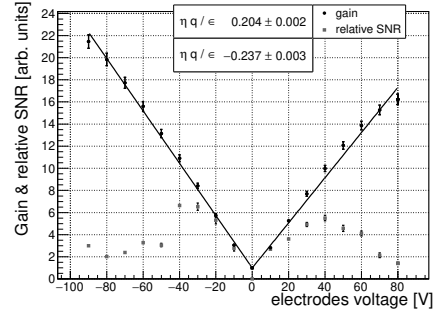}}\par
\subfloat[\label{Gain_SNR:c}][NTLLD2 gain and signal-to-noise ratio.]
{\includegraphics[width=.45\linewidth]{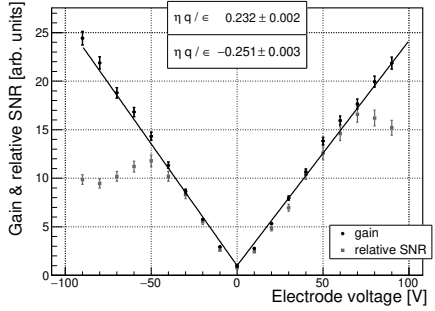}}\hfill
\subfloat[\label{Gain_SNR:d}][NTLLD3 gain and signal-to-noise ratio.]
{\includegraphics[width=.45\linewidth]{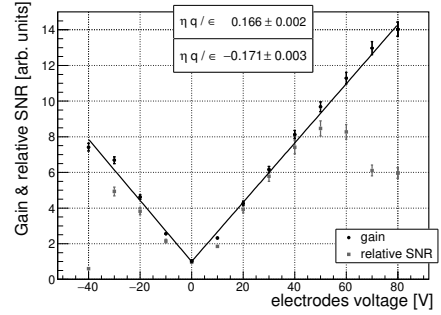}}
\caption{Examples of gain and signal-to-noise ratio as a function of the electrode bias. A linear fit is used to derive the voltage NTL gain (\emph{i.e.}, $\eta$ $\cdot$ $q$ / $\epsilon$) for LED photon  bursts of 0.85 $\mu$m wavelength photons.}
 \label{fig:Gain_SNR}
 \end{figure}
The following detector parameters have been studied: the NTL signal amplification, the signal-to-noise ratio, the signal sensitivity and the noise conditions in different environments (\emph{i.e.} different dilution refrigerators). 

The evolution of the NTL gain was measured for a given photon wavelength by injecting constant intensity LED bursts at low repetition rate (about one every 3~s) and recording the signal (amplitude of the pulses) seen by the detector, while varying the electrode voltage bias. Simultaneously, the RMS baseline noise was monitored to estimate the signal-to-noise ratio, for each bias value. The LED burst intensity was chosen to provide pulses on the detector within the linear region of the detector response, for the highest electrode bias. Fig.~\ref{fig:Gain_SNR} shows some examples of the NTL gain $G_{NTL}$ and the corresponding SNR values, as a function of electrode voltage bias. Overall, the NTL gain (slope) is almost similar with respect to the voltage polarity. Nevertheless, the data show a difference in the gain (within 10\%) when a positive/negative bias is applied. A difference with respect to electrode bias polarity is also observed, for the maximal SNR. This behaviour was already observed in the early investigations~\cite{Stark:2005,Isaila:2006}. 

Data of Fig.~\ref{fig:Gain_SNR} were fitted by a linear function to derive the value of the amplification efficiency $\eta$ (taking into account $\epsilon$ for the used light source,  Ref.~\cite{Koc:1957}). The highest $\eta$ value achieved with either the positive or the negative electrode bias is quoted in Table \ref{tab:performance}. The NTL amplification efficiency for   0.52~$\mu$m wavelength photons (which is the average wavelength of photons of the Cherenkov radiation in TeO$_2$ crystal~\cite{Casali:2017}) is evaluated from the gain at the optimal (from the point of view of the signal-to-noise ratio) electrode voltage bias. Results are summarized in Table \ref{tab:performance}. 

\begin{figure}[h!]
\subfloat[\label{results:a}][Optimal voltage for each light detectors, measured in different runs.]
{\includegraphics[width=.45\linewidth]{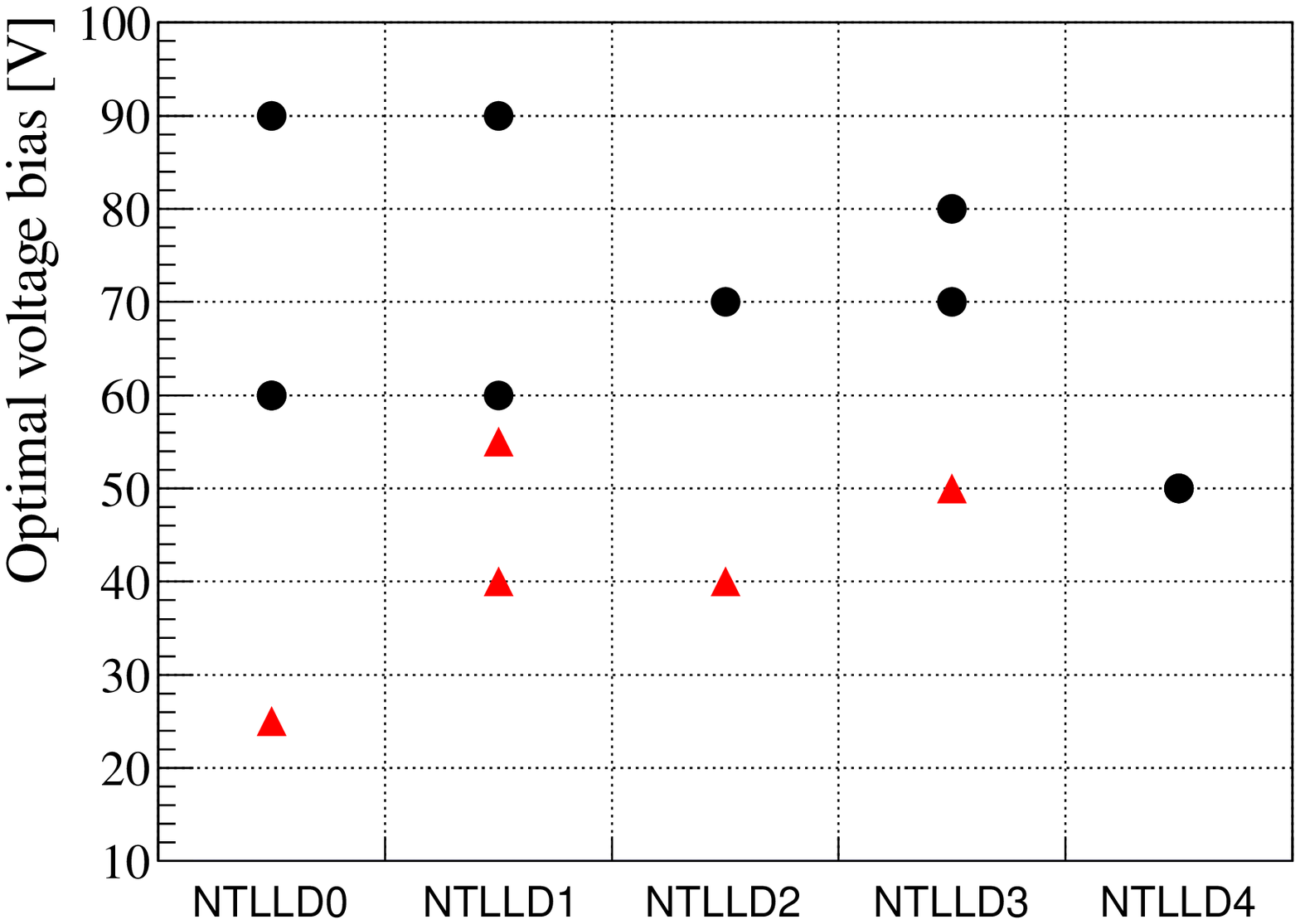}}\hfill
\subfloat[\label{results:b}][Relative SNR for each light detectors, measured in different runs.]
{\includegraphics[width=.45\linewidth]{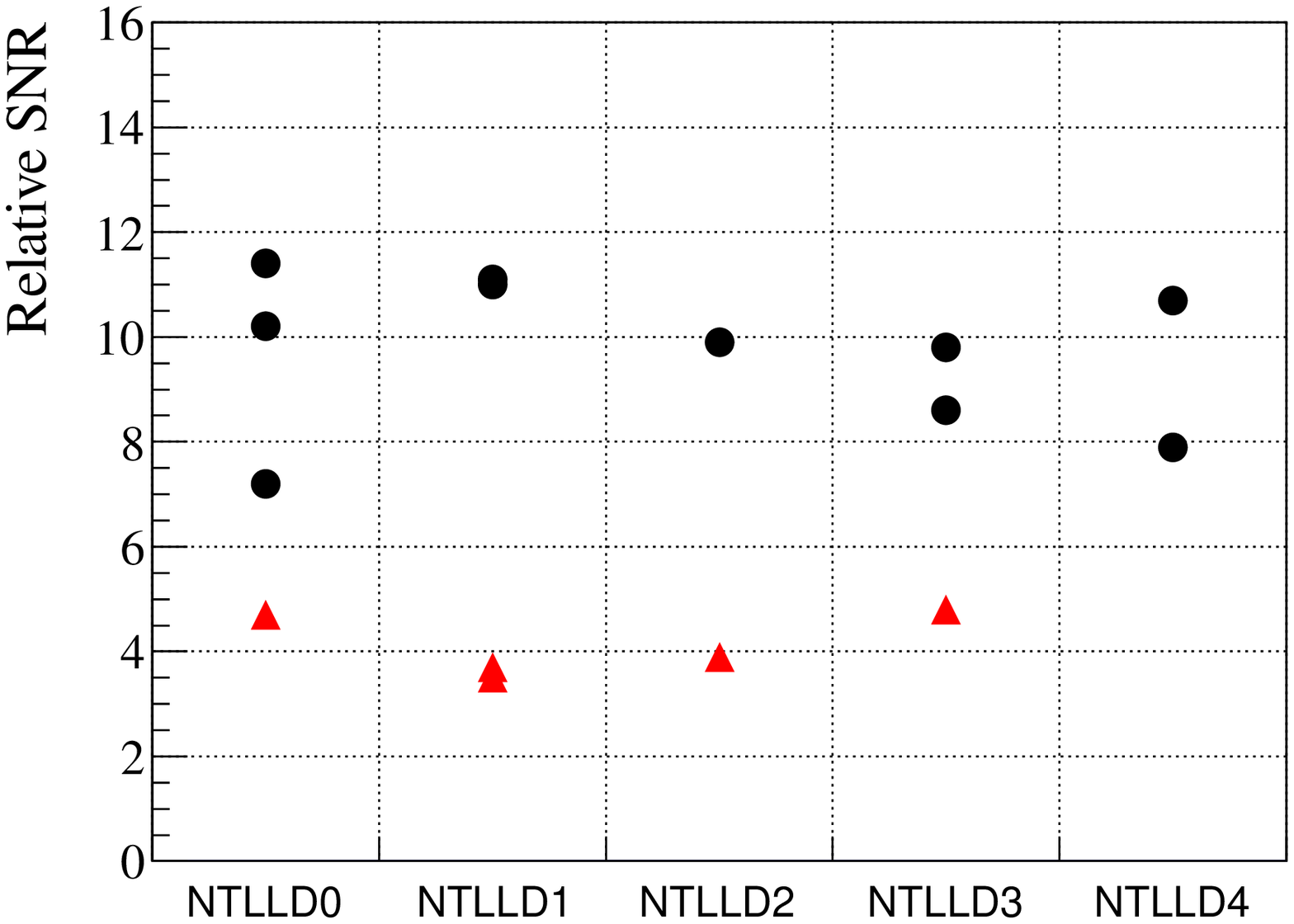}}\hfill
\subfloat[\label{results:c}][RMS noise at the optimal electrode bias as a function of the RMS noise with grounded electrodes (all detectors plotted).]
{\includegraphics[width=.45\linewidth]{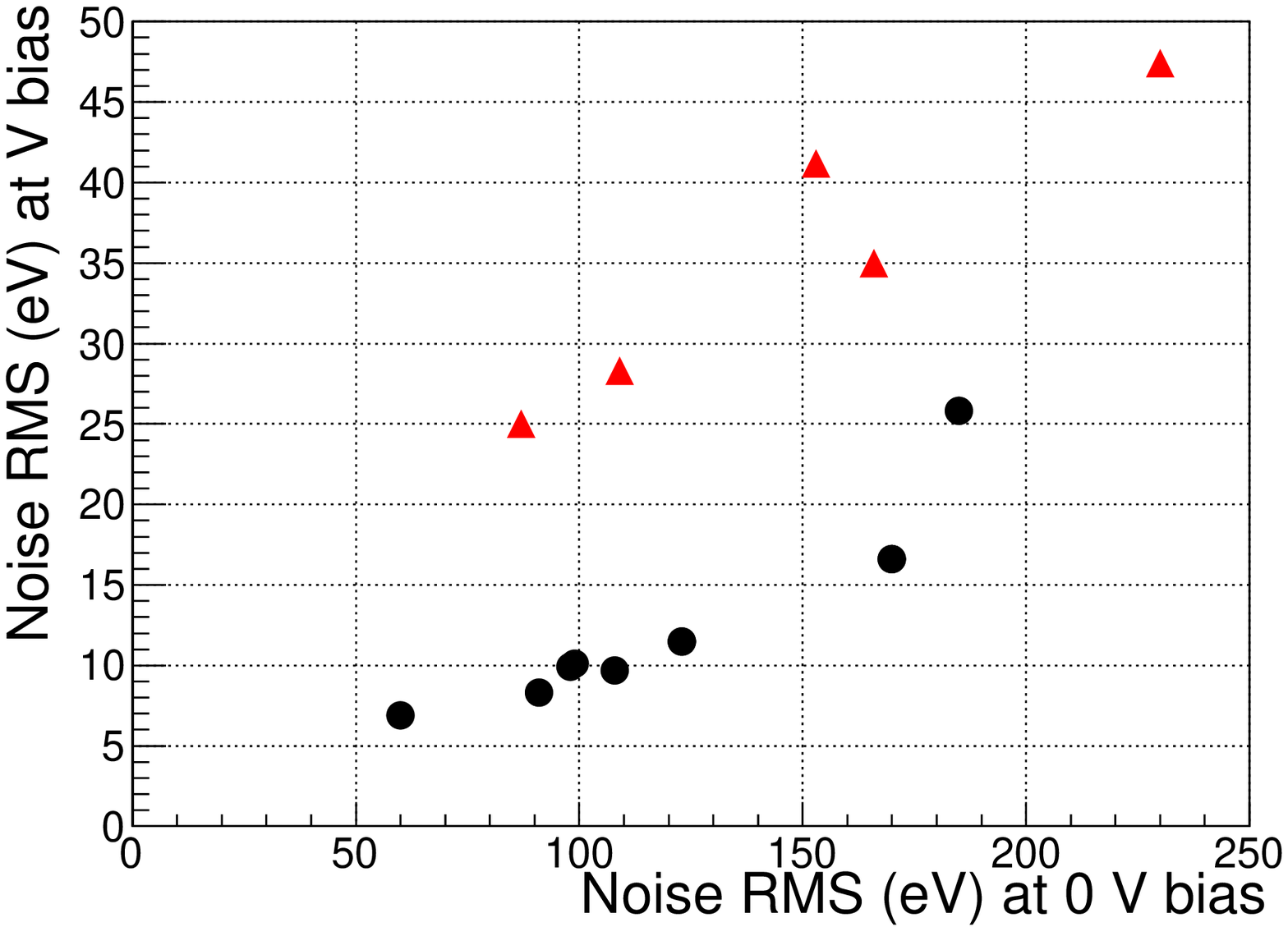}}\hfill
{\includegraphics[width=.45\linewidth]{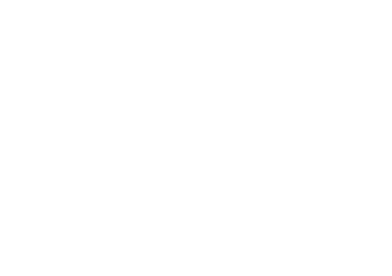}}\hfill
\caption{\textit{(Color online}) Graphical representation of some results given in Table~\ref{tab:performance}, rescaled to 0.52~$\mu$m photon wavelength. Point markers summarize the results of run-I and of those obtained in CUPID R\&D cryostat (spurious infrared radiation impinging the detectors), whereas triangle markers summarize the results obtained in run-II and EDELWEISS cryostat (radiation-tight environment).}
\label{fig:results}
\end{figure}

As previously stated, the gain of the NTL devices depends on the wavelength of the incident radiation~\cite{Koc:1957}. The measurements described in this work have been performed with different light sources: 0.85~$\mu$m and 0.95~$\mu$m wavelength photon bursts, scintillation light (peaked at $\sim$0.6~$\mu$m wavelength) and Cherenkov light ($\sim$0.52~$\mu$m wavelength, on the average).
In order to consistently compare the performances, we rescaled the relative SNR and the baseline noise, for all of them, to 0.52~$\mu$m wavelength photons.
\begin{landscape}		
\begin{table}[htb]
\centering
\scriptsize
\begin{tabular}{|l|cccc|cccc|ccc|r|}
  \hline
Detector & Set-up & Run &$T_{holder}$ & Light  							& Amplification 				& Optimal & \multicolumn{2}{c|}{Relative SNR} & $S_A$ ($\mu$V/keV) & \multicolumn{2}{c|}{Noise RMS (eV)} & Ref. \\
~				 & ~			&	ID	&[mK]& $\lambda$ ($\mu$m) 	& efficiency $\eta$   & bias V$_{el}$ (V)		& [$\lambda$] & [0.52 $\mu$m] & [V$_{el}$=0 V]	& [V$_{el}$=0 V] & [V$_{el}$] 				 & ~ \\
  \hline
NTLLD0	& A & II	& 18 & 0.85		& 0.62		& 60 		& 20.0 & 11.4 	& 0.49 	& 170 & 17 			& ~ \\   
		& ~ & ~		& ~  & 0.62		& 0.53		& ~  		& 12.5 & 10.2 	& ~ 	& ~   & ~ 			& ~ \\ 			
		& D	&	--	&n.a.& 0.52		& 0.28		& 90 		& 7.2  &  7.2 	& 1.0 	& 185 & 26 			&\cite{Pattavina:2016}\\	
		& ~	&	III	& 18 & ~		& 0.43	 	& 25 		& 4.7  &  4.7 	& 0.57 	& 166 & 35 			& \cite{Artusa:2017}\\           		
  \hline
NTLLD1	& A	& I		& 18 & 0.85		& 0.39		& 40 		&  6.6 &  3.7	& 0.74 	& 153 & 41 			& ~ \\
		& B	& --	& 20 & 0.62		& 0.48		& 90 		& 11.5 & 11.0   & 0.53  &  91 &  8			& ~ \\			
  		& D & III	& 18 & 0.52		& 0.40		& 55 		&  3.5 &  3.5	& 1.3 	&  87 & 25 			& \cite{Artusa:2017} \\
  		& E	& --	& 17 & ~		& 0.53	 	& 60 		& 11.1 & 11.1 	& 0.92 	& 108 & 10 			& \cite{Berge:2017} \\
  \hline
NTLLD2	& A & I		& 18 & 0.85		& 0.29		& 40 		&  6.8 &  3.9 	& 0.58 	& 109 & 28			& ~ \\
  		& ~ & II	& 19 & ~		& 0.41		& 70 		& 16.6 &  9.9 	& 0.83 	&  98 & 10 			& ~ \\
 \hline
NTLLD3  & A & I		& 17 & 0.85 	& 0.28		& 50 		&  8.5 &  4.8 	& 0.61 	& 230 & 47 			& ~ \\
  		& C & --	& 20 & 0.95 	& 0.33		& 80 		& 19.8 &  9.8 	& 1.1 	&  99 & 10			& ~ \\
  		& ~ & ~		& ~  & ~ 		& ~			& 70		& 17.2 &  8.6 	& ~	 	& 60$^*$ & 7$^*$ 	& ~ \\			
  \hline
NTLLD4	& A & --	& 15 & 0.85		& 0.63		& 50 		& 12.6 &  7.9   & 1.5 	& 123 & 11  		& ~ \\
  		& ~	& ~		& ~  & 0.60		& 0.52		& ~  		& 11.4	& 10.7  & ~ 	& ~   & ~			& ~ \\
  \hline

\end{tabular}
 \caption{Performance of NTL light detectors (see details in the text) characterized in different set-ups: A) Cryomech PT-405 equipped dry dilution unit (Air Liquide), B) wet dilution unit (CEA/SPEC developped), C) Cryomech PT-410 equipped dry dilution unit (Cryoconcept), D) CUPID R\&D cryostat (wet, Oxford Instrument), and E) EDELWEISS-III cryostat (semi-dry, custom made by N\'eel Institute). The run identification number (ID) is indicated only for those measurements which are common for several light detectors. The temperature of the detector holder for each measurement is indicated by $T_{holder}$.  The $\epsilon$ for the typical wavelength $\lambda$ of the used light sources is following~\cite{Koc:1957}: 1.3--1.6~eV (0.95--0.85~$\mu$m, LED), 2.2--2.3~eV (0.6--0.62~$\mu$m, scintillation) and 2.5~eV (0.52~$\mu$m, Cherenkov). The values of the amplification efficiency $|\eta|$ correspond to the NTL gain measurements with the quoted light sources. The listed electrode bias V$_{el}$ is optimal in term of the signal-to-noise ratio, relatively to the 0~V bias conditions. The best relative SNR is given for the used light sources with the quoted $\lambda$ values and the Cherenkov radiation (0.52~$\mu$m). A signal sensitivity $S_A$ is given for detectors operated without the NTL regime. The RMS baseline noise  is specified for the electrode bias equal to 0~V and V$_{el}$. The NTLLD3 baseline noise level marked with $^*$ was achieved with a pulse-tube switched off.}
\label{tab:performance}
\end{table}
\end{landscape}
A graphical representation of the results given in Table~\ref{tab:performance}, rescaled to 0.52~$\mu$m wavelength photons, is given in Fig. \ref{fig:results}.
As shown in Sec.~\ref{sec:Photonoise}, the NLT-assisted detectors are highly sensitive to the photo-current noise: a special care must be taken to shield these devices from spurious (mainly infra-red) radiation.
The plots in Fig.~\ref{fig:results} show how NTL bolometers differently behave when tested in poor (triangle markers, CUPID R\&D cryostat) and good (points,  \emph{i.e.} EDELWEISS-III cryostat) radiation-tight environment.

We can look more into detail at the performances of the NTLLD2 detector, obtained in run-I and run-II (Table~\ref{tab:performance}). The performance improvement in run-II was achieved by simply strengthening the radiation-tightness of the cryostat experimental space: the inner 50 mK copper radiation shield surrounding the detector was coated with black-velvet infrared painting~\cite{BlackVelvet}.
 The photo-current noise (Fig.~\ref{fig:results}.(a))  strongly conditions the optimal electrode bias, leading to variation of this latter as high as 100\% and a difference in the relative SNR  as large as a factor 2--3. No leakage (breakdown) currents were observed for all the 5 bolometers up electrode bias values of 90~V, which hints that, likely, the signal-to-noise ratio can be improved by further enhancing the (infrared) photon-tightness of the test environment and/or bolometer holders.

Table \ref{tab:performance} reports also the detector sensitivities and the typical noise level achieved with no NTL amplification. All detectors demonstrated sensitivities as high as $\sim$0.9~$\mu$V/keV, typical for NTD-Ge-instrumented germanium optical bolometers~\cite{Armengaud:2017,Artusa:2016}. The highest sensitivity with the NTLLD4 detector was obtained  thanks to the reduced size of the NTD-Ge thermistor. A spread by a factor of 2--3 in the baseline noise levels is observed in different set-ups (\emph{e.g.} compare the results for NTLLD1 and/or NTLLD3 reported in Table \ref{tab:performance}) and demonstrates how bolometers are fragile with respect to environmental vibrations~\cite{Olivieri:2017,Maisonobe:2018}.
\section{Discussion}
In this work we have presented a process to upgrade/improve light semiconductor bolometers, whatever the sensor technology, and enhance their performance. The process consists in the realisation of bias electrodes onto the semiconductor absorber to set an electric field within the semiconductor and drift the charge carriers created by a (ionizing) particle interaction. This  allow to benefit of the so-called NTL effect and lower the detection thresholds.

NTL-effect-based light detectors can be used to go beyond the current limit of many rare-event physics experiments.\\

In neutrinoless double-beta searches based on heat-and-light composite bolometers, low threshold light detectors are used to suppress the background. In the case of TeO$_2$ based neutrinoless double-beta decay search, no exploitable scintillation signal is to-date available to reject the alpha (dominant) background in the region of interest \cite{Brofferio:2018}. However, a particle identification can be performed via Cherenkov radiation, emitted by electrons \footnote{No light is expected for the interaction of alpha particles from natural radioactivity because of four orders of magnitude higher energy threshold required for the associated light emission ($\sim 50$~keV and $\sim 400$~MeV respectively for TeO$_2$).} \cite{Fatis:2010}. The light signal available to discriminate between the alphas and the electron interaction is of about 100~eV for 2.5~MeV deposited heat, which is within the noise level of typical doped-semiconductor-sensor light detectors. Thanks to its low threshold capability, a NTL-effect-based light detector will be able to recover the light signal, hence to suppress the background.

The next-generation double-beta decay experiment CUPID (CUORE Upgrade with Particle IDentification), which is considering TeO$_2$ as a viable option to study the double-beta decay of the candidate $^{130}$Te \cite{Wang:2015b}, could eventually benefit of the NTL effect light detector technology to suppress the background up to a factor of one hundred \cite{Poda:2017a,Novati:2018}.

In dark matter searches  as CRESST~\cite{Angloher:2016a}, the NTL technology can improve the separation between the background generated by the electron recoils and the signal coming from nuclear recoils, expected for the interaction of WIMPs (weakly interactive massive particles, hypothetically constituting the dark-matter galactic halo) with the elemental composition of the target. This will lower the dark matter detection thresholds and open new possibility to explore wider WIMP mass region. Taking also into account that a spin-independent WIMP-nucleon elastic-scattering cross section is proportional to the square of the mass number of the target, NTL bolometers will improve the capability to distinguish nuclear recoils originated by WIMP scattering off light, middle or heavy nuclei in multi-target scintillation detector \cite{Angloher:2012}.

Low-threshold optical bolometers can also be exploited for the investigation of other rare processes as the study of a beta spectrum shape of 4-fold-forbidden $\beta$ decays of $^{113}$Cd and $^{115}$In \cite{Tretyak:2017} to scrutinize the value of the axial-vector coupling constant. Its value is expected to be similar to one involved in the neutrino-less double-beta decay process \cite{Suhonen:2017}. In spite of $10^{14}$--$10^{16}$ yr half-live of these rare beta decays, the induced counting rate and subsequently the probability of pile-ups in macro-bolometers containing these nuclides can be rather high, which strongly affects the precision of the spectrum reconstruction. Instead of using the macro-bolometer to trace the  beta spectrum, one can use this latter just as a scintillator and use the scintillation signal to reconstruct the beta spectrum itself. The advantage of this detection scheme is the reduction of the pile-up rate, since a light detector can have, typically, a time response 10--100 times faster that the one of a macro-bolometer. NTL-assisted bolometric light detectors will allow in this case to reconstruct beta spectra down to threshold comparable to those reached by macro-bolometers \cite{Leder:2018}.

Moreover, they can be used in $^{100}$Mo-enriched neutrinoless double-beta decay experiments, to mitigate the irreducible background of scintillating bolometers coming from the pile-ups of the two-neutrino double-beta decay events \cite{Chernyak:2012,Chernyak:2014,Chernyak:2014}.
\section{Conclusions}
Five NTL-effect-assisted germanium bolometers to detect photons of visible and near infra-red wavelength have been fabricated at CSNSM laboratory (Orsay, France) by developing a specific fabrication process for the realization of bias electrodes on high purity germanium wafer. In this work we demonstrate how this fabrication process leads to reproducible detector performances in terms of gain, optimal electrode bias, signal-to-noise ratio, signal sensitivity and baseline noise. We also show how compulsory is the shielding against spurious (infrared) radiation of the experimental space to operate the detectors in the NTL-assisted regime and fully benefit of the NTL gain. 

The detectors, when operated at 0~V electrodes bias, \emph{i.e.} with idle NTL gain, show sensitivity of 0.5--1.5~$\mu$V/keV and baseline noise of 90--230~eV (RMS), whereas they can reach a factor 10 better performances when operated in NTL regime at 50--90~V electrode bias, showing sub 10-eV baseline noise (RMS). The technology to fabricate NTL-assisted optical bolometers is currently mature to be integrated in large-scale cryogenic rare-event search experiments such CUPID \cite{Wang:2015a}, for which hundreds of reproducible, low-threshold, high signal-to-noise ratio light detectors are required, or in composite heat-and-light bolometers which exhibit tiny light yield. 
\section{Acknowledgments}

This work was partially performed in the framework of the LUMINEU project funded by the Agence Nationale de la Recherche (ANR, France; ANR-12-BS05-004-04).


\begin{thebibliography}{10}
\expandafter\ifx\csname url\endcsname\relax
  \def\url#1{\texttt{#1}}\fi
\expandafter\ifx\csname urlprefix\endcsname\relax\def\urlprefix{URL }\fi
\expandafter\ifx\csname href\endcsname\relax
  \def\href#1#2{#2} \def\path#1{#1}\fi

\bibitem{Angloher:2016a}
G.~Angloher, et~al., {Results on light dark matter particles with a
  low-threshold CRESST-II detector}, Eur. Phys. J. C 76 (2016) 25.

\bibitem{Angloher:2016}
G.~Angloher, et~al., {The COSINUS project: perspectives of a NaI scintillating
  calorimeter for dark matter search}, Eur. Phys. J. C 76 (2016) 441.

\bibitem{Alenkov:2018}
V.~Alenkov, et~al., {Technical Design Report for the AMoRE 0$\nu\beta\beta$
  Decay Search Experiment}, arXiv: 1512.05957v1.

\bibitem{Azzolini:2018}
O.~Azzolini, et~al., {CUPID-0: the first array of enriched scintillating
  bolometers for 0$\nu\beta\beta$ decay investigations}, Eur. Phys. J. C 78
  (2018) 428.

\bibitem{Armengaud:2017}
E.~Armengaud, et~al., Development of {$^{100}$Mo}-containing scintillating
  bolometers for a high-sensitivity neutrinoless double-beta decay search, Eur.
  Phys. J. C 77 (2017) 785.

\bibitem{Poda:2017}
D.~V. Poda, {$^{100}$Mo}-enriched {Li$_2$MoO$_4$} scintillating bolometers for
  $0\nu 2\beta$ decay search: from {LUMINEU} to {CUPID-0/Mo} projects, AIP
  Conf. Proc. 1894 (2017) 020017.

\bibitem{Bobin:1997}
C.~Bobin, et~al., {Alpha/gamma discrimination with a CaF$_2$(Eu) target
  bolometer optically coupled to a composite infrared bolometer}, Nucl.
  Instrum. Meth. A 386 (1997) 453.

\bibitem{Alessandrello:1998}
A.~Alessandrello, et~al., {A scintillating bolometer for experiments on double
  beta decay}, Phys. Lett. B 420 (1998) 109.

\bibitem{Meunier:1999}
P.~Meunier, et~al., {Discrimination between nuclear recoils and electron
  recoils by simultaneous detection of phonons and scintillation light}, Appl.
  Phys. Lett. 75 (1999) 1335.

\bibitem{Pirro:2006t}
S.~Pirro, et~al., Scintillating double-beta-decay bolometers, Phys. At. Nucl.
  69 (2006) 2109.

\bibitem{Pirro:2017}
S.~Pirro, P.~Mauskopf, Advances in bolometer technology for fundamental
  physics, Annu. Rev. Nucl. Part. Sci. 67 (2017) 161.

\bibitem{Poda:2017a}
D.~V. Poda, A.~Giuliani, Low background techniques in bolometers for
  double-beta decay search, Int. J. Mod. Phys. A 32 (2017) 1743012.

\bibitem{Bellini:2018}
F.~Bellini, Potentialities of the future technical improvements in the search
  of rare nuclear decays by bolometers, Int. J. Mod. Phys. A 33 (2018) 1843003.

\bibitem{Stark:2005}
M.~Stark, et~al., {Application of the Neganov-Luke effect to low-threshold
  light detectors}, Nucl. Instrum. Meth. A 545 (2005) 738.

\bibitem{Angloher:2012}
G.~Angloher, et~al., {Results from 730 kg days of the CRESST-II Dark Matter
  search}, Eur. Phys. J. C 72 (2012) 197.

\bibitem{Fatis:2010}
T.~{Tabarelli de Fatis}, {Cerenkov emission as a positive tag of double beta
  decays in bolometric experiments}, Eur. Phys. J. C 65 (2010) 359.

\bibitem{Brofferio:2018}
C.~Brofferio, S.~{Dell'Oro}, {The saga of neutrinoless double beta decay search
  with TeO$_2$ thermal detectors}, arXiv: 1801.03580.

\bibitem{Tretyak:2002}
V.~I. Tretyak, Y.~G. Zdesenko, Tables of double beta decay data --- an update,
  At. Data Nucl. Data Tables 80 (2002) 83.

\bibitem{Alduino:2017}
C.~Alduino, et~al., {First Results from CUORE: A Search for Lepton Number
  Violation via $0\ensuremath{\nu}\ensuremath{\beta}\ensuremath{\beta}$ Decay
  of $^{130}\mathrm{Te}$}, Phys. Rev. Lett. 120 (2018) 132501.

\bibitem{Wang:2015a}
G.~Wang, et~al., {CUPID}: {CUORE} ({C}ryogenic {U}nderground {O}bservatory for
  {R}are {E}vents) {U}pgrade with {P}article {ID}entification, arXiv:
  1504.03599.

\bibitem{Wang:2015b}
G.~Wang, et~al., {{R\&D towards CUPID (CUORE Upgrade with Particle
  IDentification)}}, arXiv: 1504.03612.

\bibitem{Chernyak:2012}
D.~M. Chernyak, et~al., Random coincidence of $2 \nu 2 \beta$ decay events as a
  background source in bolometric $0 \nu 2 \beta$ decay experiments, Eur. Phys.
  J. C 72 (2012) 1989.

\bibitem{Chernyak:2014}
D.~M. Chernyak, et~al., {Rejection of randomly coinciding events in ZnMoO$_4$
  scintillating bolometers}, Eur. Phys. J. C 74 (2014) 2913.

\bibitem{Chernyak:2016}
{\relax D. M}.~Chernyak, et~al., {Rejection of randomly coinciding events in
  Li$_{2}^{100}$MoO$_{4}$ scintillating bolometers using light detectors based
  on the Neganov-Luke effect}, Eur. Phys. J. C 77 (2016) 3.

\bibitem{Tretyak:2017}
V.~I. Tretyak, Beta decays in investigations and searches for rare effects,
  talk given at Int. Workshop MEDEX 2017, Prague, Czech Republic, 29 May -- 02
  July 2017.

\bibitem{Suhonen:2017}
J.~T. Suhonen, {Value of the Axial-Vector Coupling Strength in $\beta$ and
  $\beta \beta$ Decays: A Review}, Front. Phys. 5 (2017) 55.

\bibitem{Leder:2018}
A.~Leder, et~al., {Measurement of Quenched Axial Vector Coupling Constant in
  In-115 Beta Decay and its Impact on Future $0\nu\beta\beta$ Searches}, poster
  presented at the XXVIII International Conference on Neutrino Physics and
  Astrophysics (Neutrino 2018), Heidelberg, Germany, 4--9 June 2018.

\bibitem{Neganov:1985}
B.~Neganov, V.~Trofimov, {USSR} patent no 1037771, Otkrytia i Izobreteniya 146
  (1985) 215.

\bibitem{Luke:1988}
{\relax P. N}.~Luke, Voltage-assisted calorimetric ionization detector, J.
  Appl. Phys. 64 (1988) 6858.

\bibitem{Hehn:2016}
L.~Hehn, et~al., {Improved EDELWEISS-III sensitivity for low-mass WIMPs using a
  profile likelihood approach}, Eur. Phys. J. C 76 (2016) 548.

\bibitem{Agnese:2016}
R.~Agnese, et~al., {New Results from the Search for Low-Mass Weakly Interacting
  Massive Particles with the Low Ionization Threshold Experiment}, Phys. Rev.
  Lett. 116 (2016) 071301.

\bibitem{Isaila:2006}
C.~Isaila, et~al., {Scintillation light detectors with Neganov-Luke
  amplification}, Nucl. Instrum. Meth. A 559 (2006) 399.

\bibitem{Isaila:2012}
C.~Isaila, et~al., Low-temperature light detectors: {N}eganov-{L}uke
  amplification and calibration, Phys. Lett. B 716 (2012) 160.

\bibitem{Willers:2015}
M.~Willers, et~al., {Neganov-Luke amplified cryogenic light detectors for the
  background discrimination in neutrinoless double beta decay search with
  TeO$_{2}$ bolometers}, JINST 10 (2015) P03003.

\bibitem{Defay:2016}
X.~Defay, et~al., {Cryogenic Silicon Detectors with Implanted Contacts for the
  Detection of Visible Photons Using the Neganov-Trofimov-Luke Effect}, J. Low
  Temp. Phys. 184 (2016) 274.

\bibitem{Biassoni:2015}
M.~Biassoni, et~al., {Large area Si low-temperature light detectors with
  Neganov-Luke effect}, Eur. Phys. J. C 75 (2015) 480.

\bibitem{Gironi:2016}
L.~Gironi, et~al., {Cerenkov light identification with Si low-temperature
  detectors with sensitivity enhanced by the Neganov-Luke effect}, Phys. Rev. C
  94 (2016) 054608.

\bibitem{Pattavina:2016}
L.~Pattavina, et~al., {Background suppression in massive TeO$_{2}$ bolometers
  with Neganov-Luke amplified light detectors}, J. Low Temp. Phys. 184 (2016)
  286.

\bibitem{Artusa:2017}
{\relax D. R}.~Artusa, et~al., {Enriched TeO$_{2}$ bolometers with active
  particle discrimination: Towards the CUPID experiment}, Phys. Lett. B 767
  (2017) 321.

\bibitem{Berge:2017}
L.~Berg\'e, et~al., {Complete event-by-event
  $\ensuremath{\alpha}/\ensuremath{\gamma}(\ensuremath{\beta})$ separation in a
  full-size ${\mathrm{TeO}}_{2}$ CUORE bolometer by Neganov-Luke-magnified
  light detection}, Phys. Rev. C 97 (2018) 032501(R).

\bibitem{Tenconi:2015}
M.~Tenconi, Development of luminescent bolometers and light detectors for
  neutrinoless double beta decay search, Ph.D. thesis, Universit\'e Paris-Sud
  (2015).

\bibitem{Mancuso:2016}
M.~Mancuso, Development and optimization of scintillating bolometers and
  innovative light detectors for a pilot underground experiment on neutrinoless
  double beta decay, Ph.D. thesis, Universit\'e Paris-Sud (2016).

\bibitem{Novati:2018}
V.~Novati, {Sensitivity enhancement of the CUORE experiment via the development
  of Cherenkov hybrid TeO$_{2}$ bolometers}, Ph.D. thesis, Universit\'e
  Paris-Saclay (2018).

\bibitem{Shutt:2000}
T.~Shutt, et~al., A solution to the dead-layer problem in ionization and
  phonon-based dark matter detectors, Nucl. Instrum. Meth. A 444 (2000) 340.

\bibitem{Mancuso:2014}
M.~Mancuso, et~al., An experimental study of antireflective coatings in {G}e
  light detectors for scintillating bolometers, EPJ Web Conf. 65 (2014) 04003.

\bibitem{Mancuso:2014a}
M.~Mancuso, et~al., An aboveground pulse-tube-based bolometric test facility
  for the validation of the {LUMINEU} {Z}n{M}o{O}$_{4}$ crystals, J. Low Temp.
  Phys. 176 (2014) 571.

\bibitem{Olivieri:2017}
E.~Olivieri, et~al., {Vibrations on pulse tube based Dry Dilution Refrigerators
  for low noise measurements}, Nucl. Instrum. Meth. A 858 (2017) 73.

\bibitem{Pirro:2006}
S.~Pirro, Further developments in mechanical decoupling of large thermal
  detectors, Nucl. Instrum. Meth. A 559 (2006) 672.

\bibitem{Lee:2017}
C.~Lee, et~al., Vibration isolation system for cryogenic phonon-scintillation
  calorimeters, JINST 12 (2017) C02057.

\bibitem{Maisonobe:2018}
R.~Maisonobe, et~al., Vibration decoupling system for massive bolometers in dry
  cryostats, JINST 13 (2018) T08009.

\bibitem{Olivieri:2009}
E.~Olivieri, et~al., {Space-and-surface charge neutralization of cryogenic Ge
  detectors using infrared LEDs}, AIP Conf. Proc 1185 (2009) 310.

\bibitem{Alessandrello:2000}
A.~Alessandrello, et~al., A programmable front-end system for arrays of
  bolometers, Nucl. Instrum. Meth. A 444 (2000) 111.

\bibitem{Landau:1944}
L.~D. Landau, {On the energy loss of fast particles by ionization}, J. Phys.
  (USSR) 8 (1944) 201.

\bibitem{Koc:1957}
S.~Koc, The quantum efficiency of the photo-electric effect in germanium for
  the 0.3--2 $\mu$ wavelength region, Czechosl. J. Phys. 7 (1957) 91.

\bibitem{Casali:2017}
N.~Casali, {Model for the Cherenkov light emission of TeO$_2$ cryogenic
  calorimeters}, Astropart. Phys. 91 (2017) 44.

\bibitem{BlackVelvet}
Mankiewicz NEXTEL$^{\textregistered}$~3M,~velvet-coating 811-21 and hardener
  5524.

\bibitem{Artusa:2016}
{\relax D. R}.~Artusa, et~al., {First array of enriched Zn$^{82}$Se bolometers
  to search for double beta decay}, Eur. Phys. J. C 76 (2016) 364.

\end{thebibliography}

\end{document}